\documentclass[runningheads]{llncs}
\usepackage[T1]{fontenc}
\usepackage{graphicx}
\usepackage [autostyle]{csquotes}

\begin{document}
\title{Low Total Fertility in Simple Economic Systems}
%
%
\subtitle{The Dangerous Allure of Infertility}
\author{John Stevenson\orcidID{0000-0001-8518-9997} }
%
\authorrunning{J. Stevenson}
%
\institute{Long Beach Institute, Long Beach, NY 11561 \\
\email{jcs@alumni.caltech.edu}}
%
\maketitle              
\begin{abstract}
	
Low total fertility rates throughout the world have lead to concerns about economic growth, military security, international political power, environment impacts, and quality of life. Overall total fertility rates of today's societies are complex emergent functions of culture, biology, and economic policies that are notoriously difficult to forecast.
In order to study the dynamic, stochastic nature of total fertility rates, population and wealth trajectories as functions of infertility and birth cost are generated from a minimal, endogenous, agent-based model of a simple foraging economy.
A harvesting model from mathematical ecology is added to reflect death by \textquote{natural causes}.
With these added limits of finite lifespans, decreasing total fertility rates are shown to lead to population levels consistently below the actual carry capacity of the landscape.
These below carry-capacity population levels generate higher total and per capita wealth. 
The stochastic population trajectories generated demonstrate instabilities that significantly increase the likelihood of extinction within reasonable time frames.  
Society may possibly be encouraged by this increasing wealth (and perhaps reduced environmental degradation) to continue decreasing total fertility rates, further increasing the extinction risk.
Conversely, the additional wealth might increase total fertility rates through relatively lower birth costs.
Tax-funded subsidies are added to the model to determine if directly reducing birth costs can significantly increase total fertility rates to escape these stochastic instabilities.
This research demonstrates that understanding attempts to mitigate the consequences of declining total fertility rates must include modeling of the dynamic and stochastic nature of these population trajectories.

\end{abstract}
\section{Introduction}
Population declines due to low total fertility rates (TFR) has been a fact of life for many modern societies for decades. Over 30 countries now all have TFRs below the replenishment rate \cite{ogawa2005japan,sobotka2008diverse,birg2002demographic,bongaarts2015global,mcdonald2006low}. Low TFRs also occurred in many Western countries between 1920 and 1940 and has been argued to be \textquote{an obstinate feature of modern society} \cite{van2010subreplacement}. Businesses in general assume that population decline is bad for growth \cite{birg2002demographic,guzmancecilia,coleman2011s}. The empirical literature, however, suggests that the size and growth of a population do not reduce per capita GDP nor economic growth  \cite{barroj,sheehey1996growing,coleman2011s}. The expected increases in productivity due to technology reinforce these findings \cite{cohen1995many}. Governments fear population decline for its assumed negative effects on tax revenue, military security, and international political power \cite{jones2009ultra,coulmas2007population,bainbridge2009demographic,kagan2004paradise,mcdonald2006low}.  On the other hand, a number of organizations propose population reduction to address environmental and economic crises \cite{bradshaw2014human,guzmancecilia,russell2000population}. The complexity of population dynamics thwarts any definitive forecast of TFR, determination of causation, or effects of policy changes \cite{lutz2006low,turchin2009long}.

Conversely, a minimal model of a system \cite{roughgarden}, in this case an agent-based, endogenous model of a simple foraging economy, provides repeatable, quantifiable, and stochastic explanations of relevant population and wealth dynamics \cite{stevenson2021population,stevenson2023local,stevenson2024dist}. 
This research demonstrates a dangerous coupling of population decline with increasing total and per capita wealth, potentially luring society into stochastic regimes of population dynamics that would inevitably lead to extinction in reasonable time frames. This \textquote{low fertility trap} is a widely discussed and debated demographic hypothesis  \cite{lutz2006low}.

First the details of this foraging model are provided. 
The mathematics of single species population dynamics as relevant to the foraging model are presented with the intrinsic growth rate as a proxy for TFR.
Support in the literature for the usefulness and validity of this simple, minimal model for a foraging economy is provided.
Various death by \textquote{natural causes} models are presented and compared with census data and a standard demographic hazard model.
The intrinsic growth rate declines are shown to be functions of the infertility and birth costs model parameters. 
The evolutionary selection pressures on the relevant model parameters are explored.
The stochastic population and surplus wealth trajectories emerging from the foraging model for increasing infertility and birth costs are presented and discussed.
Policy interventions through taxes to reduce birth costs are shown to significantly mitigate declining populations levels and to a lesser extent volatilities.


\begin{table}[h!]
	\begin{center}
					\begin{tabular}{|c|c|c|c|} 
						\hline
						Agent & Symbol & Value & Units  \\ 
						\hline
						vision & $v$ &  6  &  cells  \\
						movement & -- &  6 &  cells per cycle  \\
						metabolism & $m$ & 3 & resources per cycle \\
						birth cost & $bc$ & 1-900 & resources  \\
						infertility & $f$ & 1-900 & 1/probability \\
						puberty & $p$ & 1 & cycle \\
						surplus & $S$ & 0+ & resources  \\
						mutation & $\mu$ & 0-1 & probability  \\
						sociality & $ix$ & 0-2 & true/false/NA  \\
						local sharing & $shr$ & 0-2 & none/share all/share within id \\
						id & $id$ & 0-9 & group identification \\
						finite lifespan, minimum & FD & 100-1000 & cycles  \\
						finite lifespan, range & FR & 100-1000 & cycles  \\
						\hline
					\end{tabular}
				\bigbreak
						\begin{tabular}{|c|c|c|c|} 
						\hline
							Landscape & Symbol & Value & Units\\ 
						\hline
							rows & $row$ & 50 & cells \\
							columns & $col$ & 50 & cells\\
							max capacity &$R$ & 4 & resource per cell\\
							growth & $g$ & 1 & resource per cycle per cell \\
							initial & $R_{0}$ & 4 & resource, all cells\\
							\hline
						\end{tabular}
		\caption{Agent and Landscape Parameters of the ABM}
		\label{table:appA}
				\end{center}
			\end{table}

\section{Methods}

A spatiotemporal, multi-agent-based model based on Epstein and Axtell's classic Sugarscape \cite{axtell,stevenson} is used to model a simple foraging economy \cite{stevenson2021population,stevensonEcon}. As a minimum model of a system \cite{roughgarden}, the model does not attempt to calibrate to an empirical objective function. Rather, a population of agents endogenously evolves under selection pressures.
The foraging resources are evenly distributed across the landscape giving equal opportunity to all. The capabilities of each agent are identical. 
Table \ref{table:appA} provides the definition of the agents' and landscape's parameters used for this investigation. The two dimensional landscape wraps around the edges (often likened to a torus).  Vision and movement are along rows and columns only. Agents are selected for action in random order each cycle.  The selected agent moves to the closest visible cell with the most resources with ties resolved randomly.
A control volume analysis of the resource cycle was performed. Resources on the landscape grow at $g$ (1 resource per cycle per landscape cell) to a maximum of $R$ (4 resources). Resources are destroyed by the agent's metabolism at a rate of $m$ (3 resources per agent per cycle), by the cost of reproduction $bc$ which are lost (not transferred to the child), and through the death of an agent with a surplus (without an heir or in the no-inherit configuration).

\section{Population Dynamics}
The mathematical theory of continuous and discrete models of single species population dynamics are reviewed. The emergence of complexity regimes in the discrete model that are absent in the continuous model are highlighted. The challenges introduced by including finite lifespans are discussed and compared with harvesting models of mathematical ecology. The emergent population and surplus trajectories of these simulations are compared with the appropriate biological, ecological, and genetic population models. The actual and target age structures in the simulation are compared with the basic Gompertz demographic model as well as actual census data.

\subsection{Simple Models without Age Structure}
A continuous model of a single species population $N(t)$ was proposed by Verhulst (1838) \cite{verh,murray,kot} :

\begin{equation}
	\frac{dN(t)}{dt}=rN(1-\frac{N}{K}) \label{contInV}
\end{equation}
where $K$ is the steady state carry capacity, $t$ is time, and $r$ is the intrinsic rate of growth. 
This model represents self-limiting, logistic growth of a population and is used to estimate the intrinsic growth rates $r$ of the population trajectories generated by this simple model.
As such, death only occurs by starvation and as there is no other limit on lifespan, agents may sometimes live to fantastical ages generating immense fortunes (\textquote{founder's effect} \cite{stevensonEcon}).

The discrete form of the Verhust is derived from Eq.(\ref{contInV}) by replacing the derivative $dN(t)/dt$ with the difference form with a time step of $1$ \cite{murray}:
\begin{equation}
	N(t+1)=[1+r-\frac{N(t-\tau)}{K}]N(t) \label{discretV}
\end{equation}

This discrete form of the Verhulst process incorporates an explicit time delay $\tau$ in the self-limiting term as proposed by Hutchinson \cite{hutch} to account for delays seen, for example, in animal populations. With a  $\tau = 3$ to match the delay in replenishing the foraging landscape, this equation captures the steady state, oscillating, and chaotic populations trajectories which are also seen in the simple foraging model with similar intrinsic growth rates. This discrete-delayed logistic equation (Eq. \ref{discretV}) is often referred to as the Hutchinson-Wright process \cite{kot}.

\begin{figure}
	\begin{center}		
		\resizebox{\columnwidth}{!}{	
			\includegraphics[angle=-90,scale=1.0]{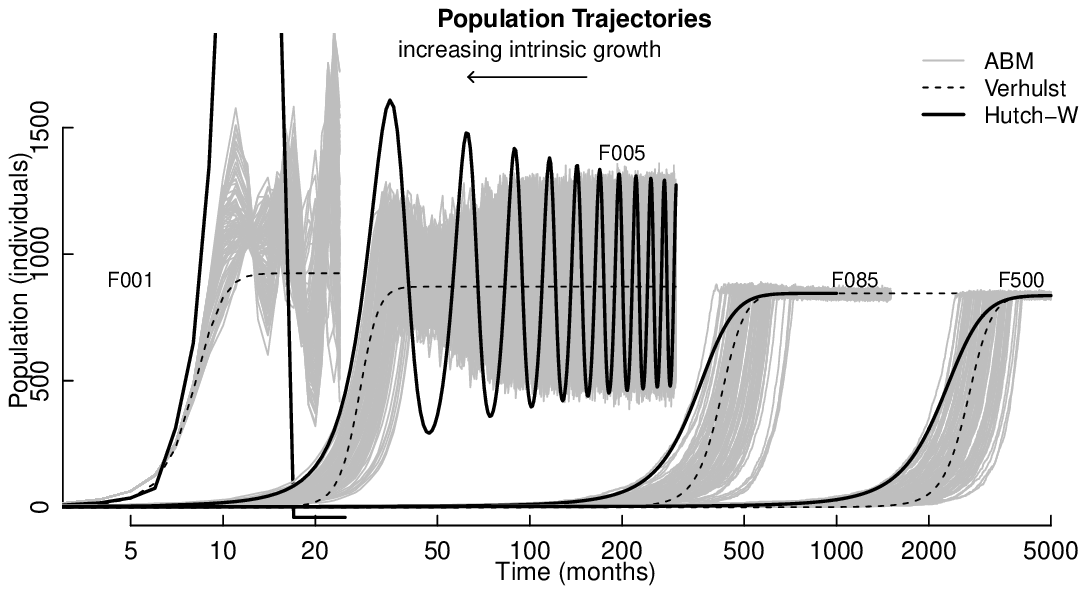}
		}
	\end{center}
	\caption{ Stochastic population trajectories showing the chaotic, oscillatory, and stable regimes for 20 differently seeded runs of the foraging model for each critical infertility. The labels 'Fnnn' refer to the infertility value (nnn) in the foraging model. The continuous Verhulst process of Eq.(\ref{contInV}) and the discrete process of Eq.(\ref*{discretV}) (labeled Hutch-W) are also provided for comparison and reference to the intrinsic growth rates of the Verhulst and Hutchinson-Wright processes.}	
	\label{fig:popTraj}
\end{figure}

Figure \ref{fig:popTraj} provides sample population trajectories for the foraging model with implied intrinsic growth rates and continuous Verhulst and discreet Hutchinson-Wright process trajectories at critical intrinsic growth rates. 
The dynamics that emerge from this simple underlying model have been shown to agree with discrete Hutchinson-Wright time delayed logistic growth models of mathematical biology and ecology \cite{murray,stevenson,hutch1,kot,stevensonEcon}, with standard Wright-Fisher class, discrete, stochastic, gene-frequency models of mathematical population genetics for finite populations \cite{ewens,moran,cannings,stevenson}, with modern coexistence theory for multiple species \cite{chessMech,stevensonX}, and with empirical income distributions of modern economies \cite{stevenson2024dist}. 
Dynamics of complex adaptive systems emerge with both intra-group and inter-group evolutionary optimizations \cite{wilsonCAS}.

Evolutionary pressure for a single group in a constant environment drives the intrinsic rate of growth into the oscillatory and chaotic regimes (to the left in Figure \ref{fig:popTraj}).

While \textquote{tragedy of the commons} extinctions occur in high-intrinsic-growth chaotic and oscillatory regimes \cite{hardin1968tragedy,ostrom1990governing}, more germane to this research are regimes of low intrinsic growth rates. Figure \ref{fig:popTraj} also shows these stable regimes and initial trajectories as driven by intrinsic growth rate.
\begin{figure}
	\begin{center}		
		\resizebox{\columnwidth}{!}{	
			\includegraphics[angle=-90,scale=1.0]{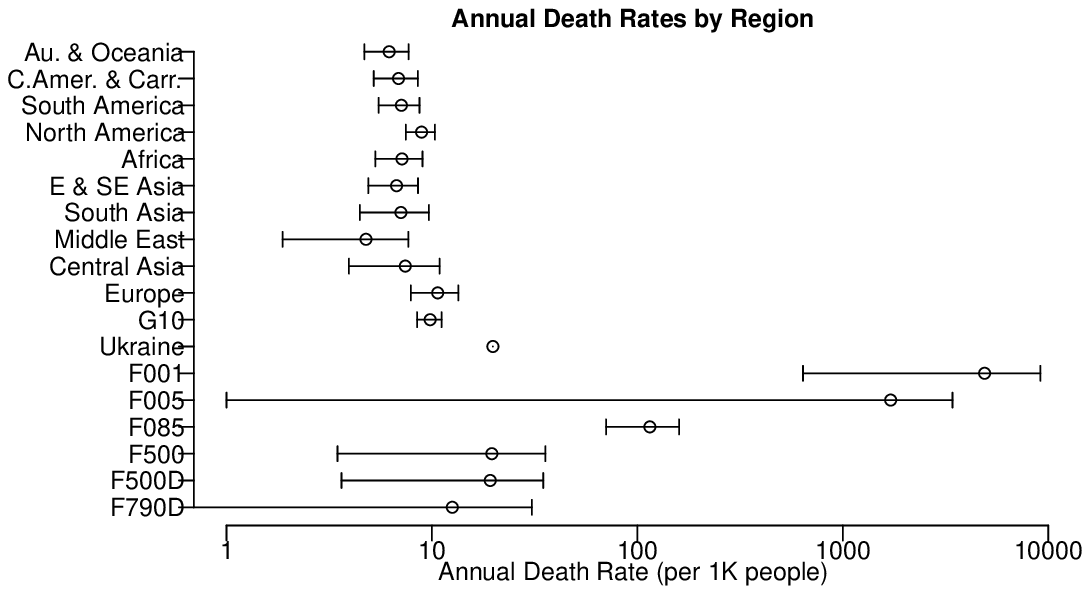}
		}
	\end{center}
	\caption{ Annualized death rates for geographic regions of the world  \cite{ciaDeathR} and for the model with various values of infertility. A relatively high infertility of F500 was necessary to bring the simulations' death rates in line with these data on human societies. F500D and F790D refer to age-structured configurations.}	
	\label{fig:deathRates}
\end{figure} 
The intrinsic growth rate of the model is modulated by infertility, puberty, and birth-cost configuration parameters as shown in Table (\ref{table:appA}). All three parameters have a similar effect on intrinsic rate of growth. The following results are based on varying inferitilty and birth cost with puberty held at 1 cycle. Given the birth cost and free space constraints are met, the probability of reproduction, $p_{f}$, is expressed as infertility $ f = \frac{1}{p_{f}}$ and labeled as F$p_{f}$ (e.g. F010 has a 10\% probability of reproduction if all other criteria are meet). Though the chaotic (F001), oscillatory (F005), and stable (F085) regimes are of great interest (Figure \ref{fig:popTraj}), they do not reflect the death rates of current human societies as shown by Figure \ref{fig:deathRates} \cite{ciaDeathR}. An increase to infertility F500 achieved death rates comparable with modern human societies with "annual" defined as 10 action cycles for the simulations. Figure \ref{fig:popTraj} also highlights the stable population levels at carry capacity  \cite{murray,kot,cohen1995many}.

		
		
		

\subsection{Age Structure Modeling of Death by Natural Causes}
In a single species population with no age structure, the only cause of death is by starvation and any positive intrinsic growth will eventually drive the population to the environment's carry capacity $K$, which, for the foraging model, is given by
		\begin{equation}
			K=g \cdot row \cdot col/m
		\end{equation}
where $g$ is the resource growth per landscape cell per cycle, $row$ and $col$ are the rows and columns of the cells respectively, and $m$ is the metabolism of each agent (resources/cycle).
		
Incorporation of an age structure with death possible by \textquote{natural causes} is accomplished with finite lifespan modeling (FL). The theoretical modeling becomes more complex and dependent on whatever external age structure is chosen. 
These continuous theoretical models (e.g. Von Foerster Equation \cite{kot,murray}) are beyond the scope of this paper. A common demographic model is the Gompertz distribution \cite{gompertz1825xxiv,missov2013gamma}, used in mortality modeling, which provides an heuristic for the distribution of ages of death based on an exponentially increasing hazard rate. The probability density function $d(t)$ for death at age $t$ is given as
\begin{equation}
	d(t) = b \exp(a t) \exp\left(-\frac{b}{a} (\exp(a t) - 1)\right)
	\label{eq:gomp} 
\end{equation}
where $t$ is the agent's age, $a$ is the rate of increase in mortality, and $b$ is the baseline hazard . These Gompertz distributions can be reasonably fit to skewed Gaussian distributions for computational efficiency.  

A guiding factor in building a minimal model of a system is to not assume any specific model or heuristics for behaviors of the agents. For example, though single species logistic growth model distributions emerge from population dynamics of this simple model (Figure \ref{fig:popTraj}), the underlying agent behaviors have no association with that modeling. The FL model chosen for inclusion of age structure and the resultant deaths by natural is straightforward, simple, and has precedent in harvesting models for renewable populations. Agents' ages are limited to human lifespans with 10 action cycles as a year. Agents' age of death due to \textquote{natural causes} are generated by a finite lifespan heuristic with a flat probability of death starting at age allele $FL_{min}$ and ranging for allele $FL_{range}$ years. The discrete population process in the foraging model can now be represented from Eq.(\ref*{discretV}) as
		
		\begin{equation}
				N(t+1) =[1+r-\frac{N(t-\tau)}{K}]N(t) - \sum_{i=1}^{N} H(a(N(t)_{i})>FL_{min})/(FL_{range})
			\label{eq:ageS}
		\end{equation}
where $H$ is the heavyside function with a value of 1 for those population members $N_{i}$ whose age $a$ is greater than minimum age boundary $FL_{min}$ and 0 otherwise.
		
Defining $N(t)_{old} = \sum_{i=1}^{N} H(a(N(t)_{i})>FL_{min})$ as all members of the population at time step $t$ whose age is above $FL_{min}$, and $p_{old}$ as the probability that an aged individual will die of natural causes in this $t$ time step as $p_{old} = 1/FL_{range}$, Eq. \ref{eq:ageS} can be simplified to:
		\begin{equation}
				N(t+1) =[1+r-\frac{N(t-\tau)}{K}]N(t) \\
				-N(t)_{old}\cdot p_{old}
			\label{eq:ageSsimple}
		\end{equation}		
Equation (\ref{eq:ageSsimple}) has the form of models from mathematical ecology used for determining sustainable harvesting \cite{kot,murray} with the term for death by natural causes $-N(t)_{old}\cdot p_{old}$ analogous to the yield term in these harvesting models. The expectation is a steady state carry capacity is attained that is reduced from the no-age structure carry capacity. A new lower steady state carry capacity will be achieved as long as the yield is not \textquote{excessive}. For the foraging model, the probability $p_{old}$ for sampling the target age of death is determined by the minimum and range values for the finite life heuristic. The ability to replace these \textquote{harvesting} losses is determined by the infertility and birth cost. The addition of FL had no measurable effect on death rates (see F500D in Figure \ref{fig:deathRates}), though forty percent of the deaths were due to FL rather than starvation alone. F790D in Figure \ref{fig:deathRates}  shows that the increasing the infertility even further does not change the mean death rate though it does increase its variance. 

	\begin{figure}
	\begin{center}		
		\resizebox{\columnwidth}{!}{	
			\includegraphics[angle=-90,scale=1.0]{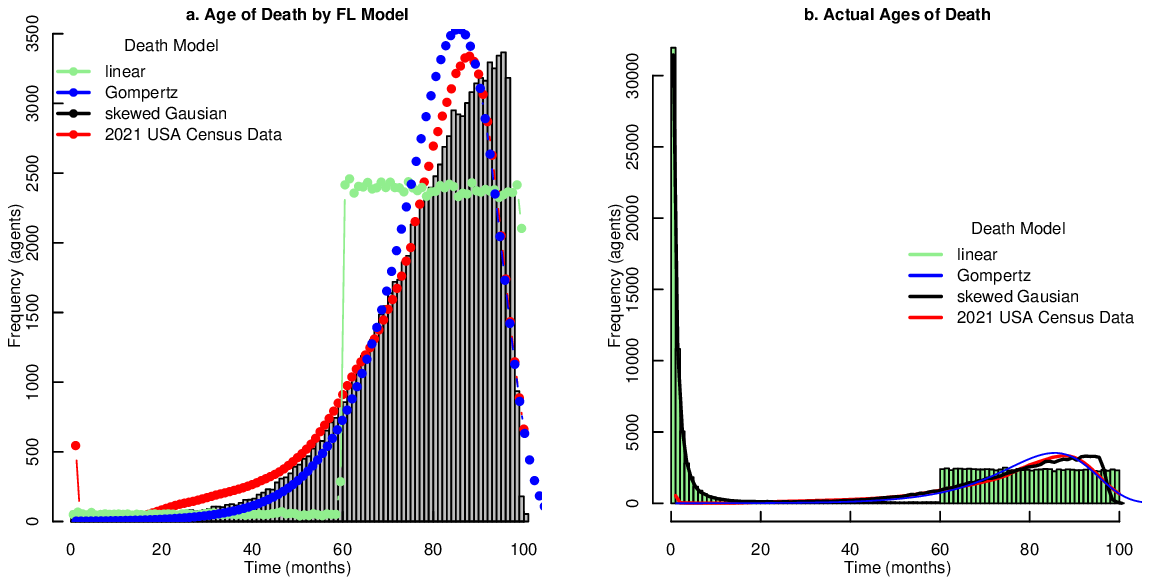}
		}
	\end{center}
	\caption{ Actual and FL models' death distributions. a) Each agent at birth samples an age of death by natural causes from the indicated FL model. The census data \cite{arias2023united} presented are actual ages of death for the referenced cohort (100,000 people). The FL model's mean frequency of ages of death over 10 differently seeded runs for a length of 1000 years is scaled to the size of the census cohort. b) The mean frequencies of ages of death over 10 differently seeded runs of the skewed Gaussian and linear FL modes over 1000 years. The actual number of deaths in the simulation exceed the census and Gompertz distributions by the approximate 50,000 actual deaths in the first three months after birth.}	
	\label{fig:deathAges}
\end{figure} 

\subsection{FL Age of Death Distributions}

Figure \ref{fig:deathAges} compares the age of death distributions for the two FL models (harvesting and the skewed Gaussian fit to the census data) with the United States Census data for 2021 \cite{arias2023united} and with the Gompertz distribution (Eq. \ref{eq:gomp}) fit to the census data. Figure \ref{fig:deathAges}a presents the linear and skewed Gaussian FL models mean sampled ages of death by natural causes over 10 differently seeded runs of length 1000 years. The total number of deaths is scaled to match the size of the census cohort (100,000 people). Figure \ref{fig:deathAges}b provides the actual ages of death in the simulations for the skewed Gaussian and linear FL models for 10 differently seeded runs lasting 1000 years. The actual mean number of deaths in the simulation exceed the census and Gompertz distributions by the approximately 50,000 actual mean deaths in the first three months.

\begin{figure}
	\begin{center}		
		\resizebox{\columnwidth}{!}{	
			\includegraphics[angle=-90,scale=1.0]{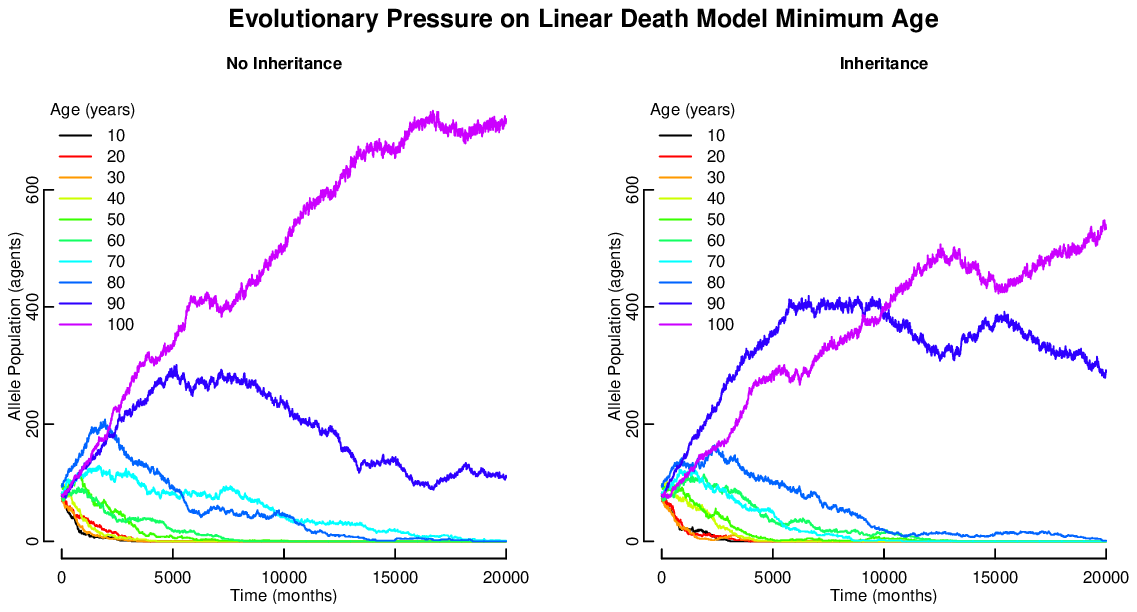}
		}
	\end{center}
	\caption{ Evolution pressures on the finite lifespan model's minimum age (FD) in a steady state environment select for the oldest possible lifespans. Models without inheritance (a) have the strongest pressure on the FD allele though, even without inheritance (b), the selection pressure on the FD allele is strong.}	
	\label{fig:evoFL}
\end{figure}

The FL model parameters ($FL_{min}$ and $FL_{range}$) were implemented as alleles in the agents chromosome and subjected to evolutionary pressure in this constant environment. As Figure \ref{fig:evoFL} shows, the expected result of selection pressure in this constant environment is to increase the lifespan as long as possible. The pressure from populations that do not inherit surplus from dying agents (Figure \ref{fig:evoFL}a) is even greater than that for environments with inheritance (Figure \ref{fig:evoFL}b). Future work will attempt to find model configurations that, no doubt under the pressure of a changing environment, would select for shorter lifespans similar to the seasonal conditions where coexisting and commensal populations of two species of different infertilities emerged from this minimal model \cite{stevensonX}.

\subsection{Surpluses for Populations with Age Structure}
		
While populations for the foraging model that attain a steady state carry capacity ensure that the resources input to the control volume match the resources exiting the control volume, the total level of the population's wealth and mean wealth is significantly affected by both the harvesting and the various inheritance options. Using the no inheritance option means that all surplus stored by an agent dying of natural causes is a sunk cost. This option seems most appropriate for a simple foraging economy. In addition, since this research is suggesting that increasing population wealth motivates reductions in fertility, the larger sunk costs of the no-inheritance option is the worst case for this argument. Another inheritance option used in these analyses is inheritance to all surviving direct descendants (children but not grandchildren). This option still has some surplus sunk costs due to a lack of living direct descendants but greatly increases the total wealth and mean surplus of the population over time. Without any surplus sunk costs (for example by selecting any random living agent to receive the inheritance), the total wealth becomes unbounded and suggests the need for a more complicated model with perhaps eventual spoilage of the resources or an inflation model. 

The model also has the capability to tax the agents' surpluses either as an income tax or as an estate tax on inheritances \cite{stevenson2024dist}. The estate tax configuration is used to generate subsidies to reduce birth costs in an attempt to mitigate the effects of low TFR.   
		
\section{Population Trajectories at Low Intrinsic Growth}
The sensitivities of population and surplus trajectories to infertility and birth cost are described. The transitions  from steady state to decaying regimes are highlighted. A tax policy to fund a birth cost subsidy is described and the effects of this subsidy are explored.

\subsection{Low Intrinsic Growth due to High Infertility}		
\begin{figure*}[h]
	\begin{center}		
			\resizebox{\textwidth}{!}{	
				\includegraphics[angle=-90,scale=1.0]{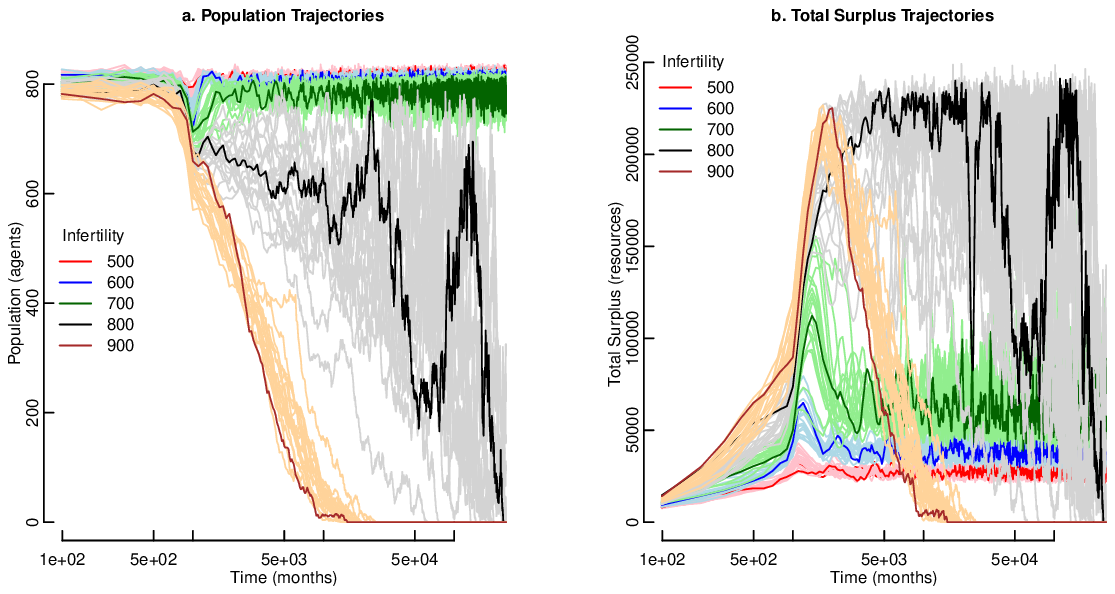}
			}
		\end{center}
		\caption{ a) Dynamic age-structured population trajectories for 20 differently seeded runs at various high values of infertility. As the infertility increases, at first the stable populations decrease slightly. At some point (F800), the dynamics transition to a decay regime with extinction inevitable.  b) Total surplus trajectories for these same 20 differently seeded runs at high values of infertility. The corresponding total surplus substantially increases with lower but stable population levels until transition to decay. At transition, at first a remarkable spike in total surplus occurs but stochastic extinction awaits, slowly for the F800 case but much more quickly for F900.}	
		\label{fig:dCrash}
	\end{figure*}

Figure \ref{fig:dCrash}a plots population trajectories of 20 randomly seeded, age-structured populations for five different values of infertility and no inheritance. As the infertility increases from the stable F500D configuration, the population at first settles at progressively lower carry capacities with higher population level variances in a similar fashion to what is seen in the harvesting models with non-excessive yield. Figure \ref{fig:dCrash}b shows the corresponding total surplus trajectories for these runs. The transition to chaotic and decaying regimes occurs around F800D, with initially a shocking rise in total surplus at a much lower peak population. Almost immediately though, the population trajectory and, more slowly, the total surplus trajectory, both decay, inevitably heading towards extinction. 
		

\subsection{Low Intrinsic Growth due to High Birth Costs}		
\begin{figure*}[h]
	\begin{center}		
			\resizebox{\textwidth}{!}{	
				\includegraphics[angle=-90,scale=1.0]{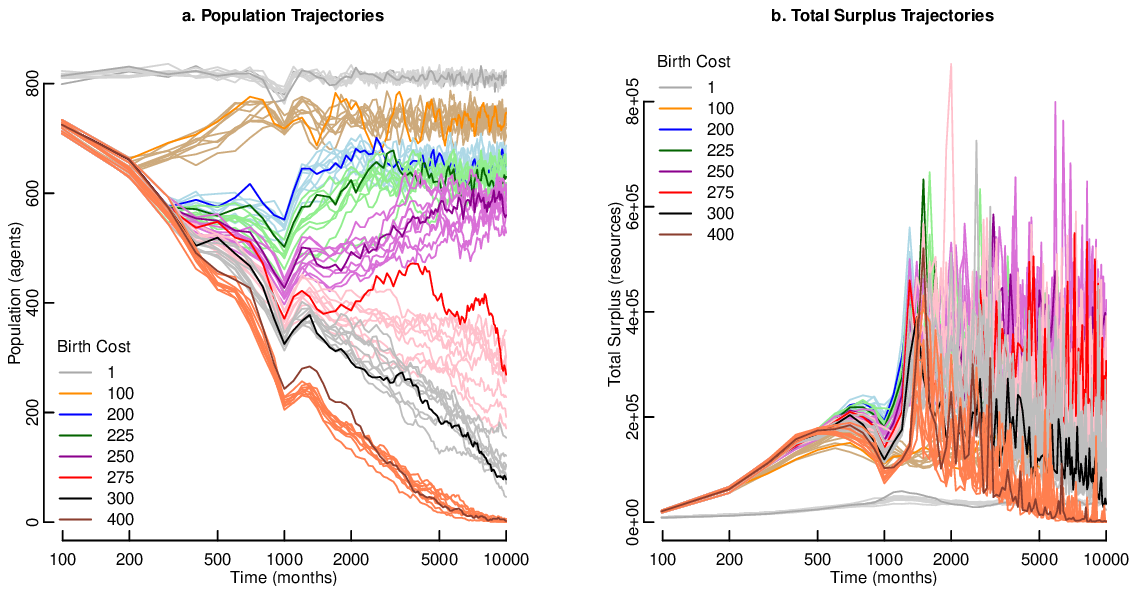}
			}
		\end{center}
		\caption{ a) Dynamic age-structured population trajectories for 10 differently seeded runs at various values of birth cost. As the birth cost increases, first populations decrease slightly below carry capacity but remain stable. At some point (birth cost between 250 and 275), the dynamics transition to a decay regime with extinction inevitable.  b) Total surplus trajectories for these same 10 differently seeded runs at the same birth cost. The corresponding total surplus substantially increases with lower but stable population levels until transition to decay. At transition, as was the case for the high infertility runs, a remarkable spike in total surplus occurs but stochastic extinction awaits, slowly for the birth cost of 300 case but much more quickly for birth cost of 400.}	
		\label{fig:dCrashBC}
	\end{figure*}

Figure \ref{fig:dCrashBC}a presents dynamic age-structured population trajectories for 10 differently seeded runs at various values of birth cost. As the birth cost increases, populations begin to decrease below carry capacity but remain stable as one would expect with a sustainable harvest. At some point (birth cost greater than 250), the dynamics transition to a decay regime with extinction inevitable.  Figure \ref{fig:dCrashBC}b tracks the total surplus trajectories for these same 10 runs at the same birth cost. The corresponding total surpluses substantially increase with lower but stable population levels until transition to decay. At transition, a remarkable spike in total surplus occurs but stochastic extinction awaits, slowly for the birth cost of 300 case but much more quickly for birth cost of 400.

\subsection{Tax Policy Mitigation of High Birth Costs}
An estate tax was chosen to fund a birth cost subsidy policy. 
While estate taxes in practice do not generate meaningful tax revenues for redistribution
\cite{bird2013death,caron2012occupy}, perhaps because they are very difficult to administer and easy to evade, they do generate significant support in the literature for reducing inequality
\cite{caron2012occupy,drometer2018wealth,hoover1927economic,aaron1992reassessing,bird2013death}.
The estate tax has also been shown for this minimal model to significantly reduce inequality even without redistribution \cite{stevenson2024dist}.
This estate tax option was implemented using the entire surplus of any agent dying to fund the Birth Cost Fund. 
For each action cycle, agents were selected in random order and if the agent qualified for reproduction based on puberty, available landscape space for the child, and the stochastic sampling of infertility, then an attempt is made to fund this agent's birth cost fully or partially from the available funds regardless of the agent's own surplus. 

\begin{figure*}[h]
	\begin{center}		
			\resizebox{\textwidth}{!}{	
				\includegraphics[angle=-90,scale=1.0]{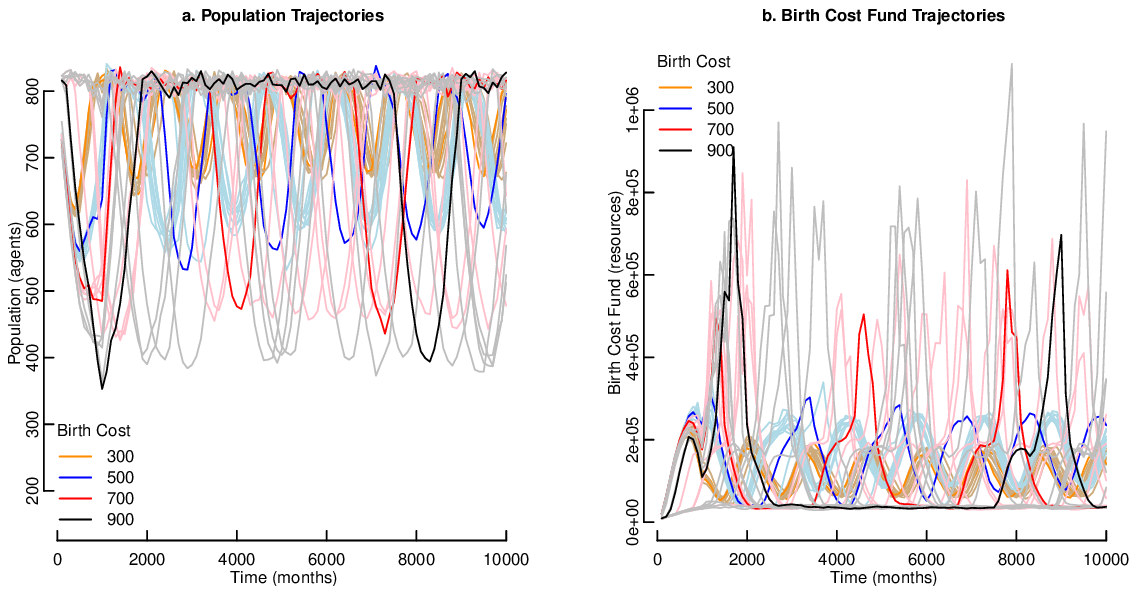}
			}
		\end{center}
		\caption{ a) Dynamic age-structured population trajectories for 10 differently seeded runs at high values of birth cost subsidized from estate taxes. As the birth cost increases, the populations mostly remain at carry capacity with large oscillations to lower levels when the subsidies are consumed. At these lower population levels, the birth cost fund is replenished and the population recovers.  b) The Birth Cost Fund trajectories for these same 10 differently seeded runs at high values of birth cost. The fund value oscillates inversely with population levels, generating higher magnitudes at lower frequencies with increasing birth costs.}
		\label{fig:dCrashSub}
	\end{figure*}

Figure \ref{fig:dCrashSub}a presents age-structured population trajectories for 10 differently seeded runs at high values of birth cost subsidized from estate taxes. As the birth cost increases, the populations mostly remain at carry capacity with large oscillations to lower levels when subsidies are not available. At these lower population levels the agents are able to easily harvest surpluses and the Birth Cost Fund is replenished enabling the population to recover.  Figure \ref{fig:dCrashSub}b plots the Birth Cost Fund level trajectories for these same 10 differently seeded runs at high values of birth cost. The fund value oscillates inversely with population level, generating higher magnitudes at lower frequencies with increasing birth costs. Despite these exceptionally high birth cost, the populations remain relatively stable.		
		
		

			
\section{Discussion}
While the economic and sociology literature presents conflicting opinions on the benefits and costs of decreasing population due to low intrinsic growth rates, these simulations show extinction risk may also arise from the stochastic dynamics of the population. When the population is significantly reduced below carry capacity due to low TFRs, the dynamic regime will transition away from stability. The surprising and frightening possibility supported by the model is that a positive feedback loop could exist between decreasing intrinsic growth rates and increasing individual wealth (\textquote{low fertility trap} \cite{lutz2006low}). With advancing productivity due to technology, the allure of increasing personal wealth may be coupled with the possibility of environmental and ecological benefits.
Growth rates have direct effects on the steady state total surplus. At very high growth rates a \textquote{tragedy of the commons} emerges, leading to periodic mass die-offs and, eventually, extinction \cite{hardin1968tragedy,ostrom1990governing}. Growth rates below replacement levels result in higher population variances and more likely demographic collapses for initial populations at carry capacity. While these decreasing intrinsic growth rates at first lead to a higher steady state mean surplus, a transition to a demographic collapse becomes an increasingly likely stochastic event \cite{escudero2004extinction}. 
In this minimal model, an estate tax policy supports a relatively stable population near the carry capacity of the landscape despite birth cost levels that would otherwise lead to a quick extinction (as well as having a side benefit of reduced inequality). While the minimal model views birth cost as a single parameter, recent literature suggests policies that address all the costs of child-rearing (medical, parental leave, child-care, housing costs) as well as equity issues (gender-gap income, work-family balance) have positive impact on total fertility rates in line these results \cite{mcdonald2006low,hubenthal2011organisation,lutz2006low}.
		
These population dynamics suggest that there is a sweet spot between the \textquote{tragedy of the commons} high intrinsic growth economies and the low intrinsic growth rate (high infertility) and harvesting (death by \textquote{natural causes}) extinction prone economies that settle at steady state population levels at or just below harvest-reduced carry capacities, perhaps with tax-based subsidies applied to reducing birth costs.

 \bibliographystyle{splncs04}
 \bibliography{/home/jack/biblio/april_24.bib} 
%
\end{document}